\documentclass[11pt]{article}

\usepackage[preprint]{acl}

\usepackage{times}
\usepackage{latexsym}

\usepackage[T1]{fontenc}

\usepackage[utf8]{inputenc}

\usepackage{microtype}

\usepackage{inconsolata}

\usepackage{graphicx}

\usepackage{times}
\usepackage{latexsym}
\usepackage[T1]{fontenc}
\usepackage[utf8]{inputenc}
\usepackage{graphicx}

\usepackage{amsmath}
\usepackage{booktabs}

\newcommand{\scorecell}[2]{\begingroup\setlength{\fboxsep}{0.3pt}\colorbox{green!#1!red!18!white}{#2}\endgroup}
\title{When Direction Matters: Diagnosing Relational Instruction Following in Dense Retrieval}

\author{Zhanyu Wu\textsuperscript{1},
  Richong Zhang\textsuperscript{1,2}\thanks{Corresponding author.} \\
  \textsuperscript{1}CCSE, School of Computer Science and Engineering, \\
  Beihang University, Beijing, China \\
  \textsuperscript{2}Zhongguancun Laboratory, Beijing, China \\
  \texttt{\{wuzy24, zhangrc\}@act.buaa.edu.cn}}

\begin{document}
\maketitle
\begin{abstract}
Dense retrievers increasingly claim the ability to condition retrieval on natural-language relevance criteria beyond a short query. We identify a specific weakness in this capability: the evaluated models perform well when lexical or attribute cues distinguish the target, but become much less reliable when relevance depends on a directional relation over the same entities. We introduce SwitchIR, a controlled diagnostic that contrasts Constraint Shift and Explicit Filtering with Relational Flip, where the query and document pair remain fixed while the requested relation direction reverses. Across representative embedding and late-interaction retrievers, the primary joint metric is high on the two cue-available controls and consistently low on Relational Flip; the same ordering persists in an LLM-assisted, manually reviewed realistic variant. An exact margin decomposition shows that both directions are correct only when the direction-sensitive query--document interaction exceeds the direction-invariant pair preference. Empirically, direction-sensitive changes are common, but their aligned interaction is usually too weak to overcome that preference. Diagnostic fine-tuning shows that models can learn relational behavior that generalizes across relation families, with stronger generalization in the larger model.
\end{abstract}

\section{Introduction}

Dense retrieval encodes queries and documents into a shared vector space, supporting efficient semantic matching over large corpora and driving progress in open-domain question answering and web search \citep{karpukhin2020dense,xiong2020approximate,izacard2021unsupervised}. As such retrievers are increasingly embedded inside retrieval-augmented generation, conversational search, and personalized assistants, the relevance target a user wants is rarely captured by the query alone: real intent often carries finer-grained criteria such as filtering conditions, audience constraints, or relation directions, requiring the retriever to adjust its relevance target according to natural language instructions.

Recent instruction-aware retrievers perform well on evaluations that vary explicit constraints \citep{weller2025followir}, user persona and context \citep{oh2024instructir}, or document-level attributes such as audience, format, and negation \citep{zhou2025beyond}. These evaluations show that current models can condition their relevance judgments on natural-language criteria, but they leave direction changes with nearly unchanged vocabulary unresolved.

\begin{figure}[!tbp]
\centering
\includegraphics[width=\linewidth]{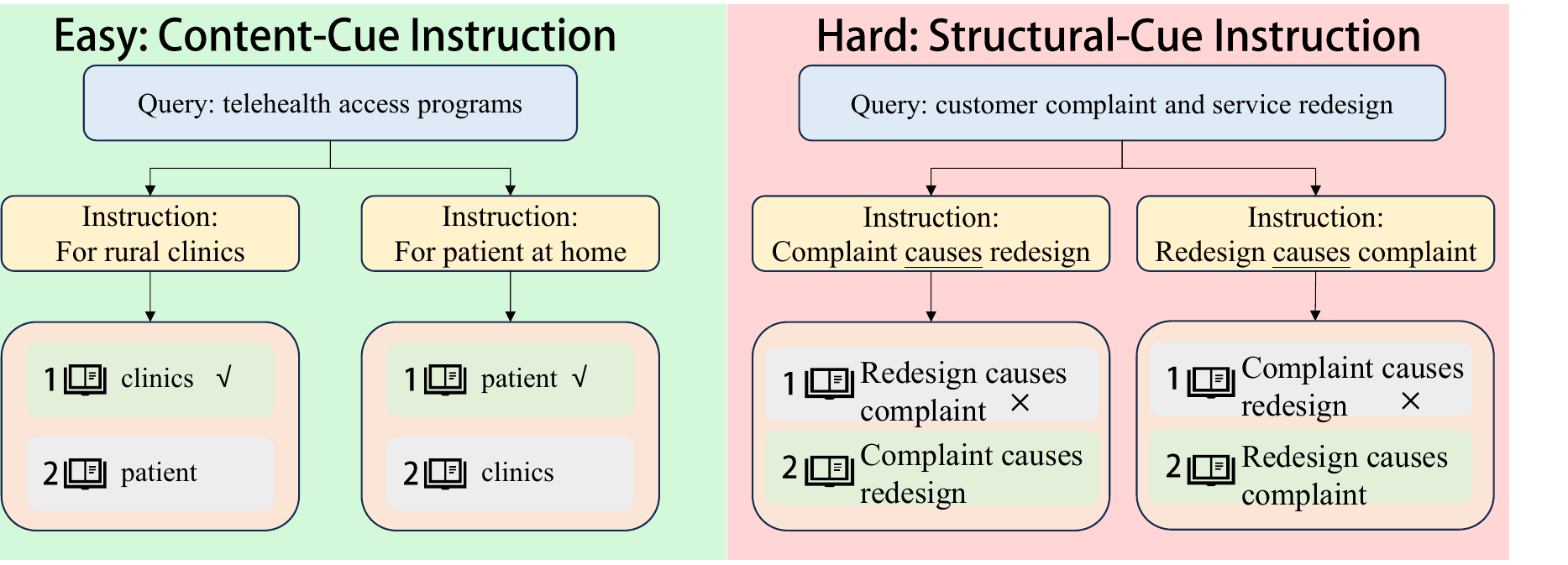}
\caption{Retrievers reliably rank the correct document when instructions differ by a distinguishing keyword (left), but often rank the reversed-relation document higher when the same words describe an opposite role assignment (right).}
\label{fig:motivation}
\end{figure}

In these evaluations, relevant and irrelevant documents can often be distinguished through explicit persona, constraint, attribute, or negation cues, leaving relational role binding largely unisolated. Many search intents, however, differ only in relation direction or semantic-role assignment while preserving the same entities and content words. Figure~\ref{fig:motivation} illustrates this diagnostic gap.

To directly evaluate whether retrievers can resolve such relational distinctions, we introduce SwitchIR, a controlled diagnostic set organized around instruction contrasts attached to a fixed query and a fixed pair of mutually hard-negative documents. Constraint Shift and Explicit Filtering test changes supported by lexical or attribute cues. Relational Flip reverses the target-defining relation (e.g., ``A causes B'' versus ``B causes A'') while holding the entities and token inventory nearly fixed, echoing broader concerns about directional relational generalization \citep{berglund2024reversal}. 

The experiments show the same category ordering on controlled and realistic data: models perform strongly on Constraint Shift, decline on Explicit Filtering, and remain weakest on Relational Flip. Our representation analysis gives an exact account of the final gap. A ranking reverses correctly in both directions only when the direction-sensitive query--document interaction exceeds the direction-invariant preference for one document in the pair. The dominant empirical failure region contains an aligned direction signal that is too weak to overcome this pair preference. Diagnostic training further shows that the directional decisions are learnable, while transfer to unseen relation families depends on training diversity and model capacity.

Our contributions are threefold. First, we formulate relational instruction following as a controlled pairwise ranking problem and introduce SwitchIR in controlled and realistic variants. Second, we establish a consistent zero-shot gap between cue-available instruction changes and relation-direction reversal across embedding, late-interaction, and reranking models. Third, we derive the exact interaction-versus-preference condition governing successful reversal and use diagnostic training to identify relation specificity and model-size-dependent transfer.

\section{Background and Related Work}

\paragraph{Dense retrieval.}
Dense retrieval encodes queries and documents into a shared embedding space and ranks candidates by vector similarity, moving beyond sparse lexical matching \citep{robertson2009probabilistic,karpukhin2020dense}. Follow-up work has strengthened this paradigm through harder negative mining, late interaction, and large-scale unsupervised pre-training \citep{xiong2020approximate,khattab2020colbert,izacard2021unsupervised}.

\paragraph{Instruction-aware embedding and retrieval.}
A growing line of work equips embedding and retrieval models with the ability to condition on natural-language instructions in addition to the query \citep{asai2023task,pan2024i3,su2023one}. One line of work builds general-purpose embedding models, scaling model size and training data to achieve strong transfer across retrieval and embedding tasks, with instruction-following emerging as one of several capabilities \citep{wang2022text,wang2024improving,muennighoff2025generative,zhang2025qwen3}. A second line of work instead targets instruction-following retrieval directly, either by training bi-encoder retrievers on instruction-augmented data with instruction negatives \citep{zhuang2025towards,zeng2026dual}, or by fusing the instruction representation into the query through an explicit attention mechanism before contrastive training \citep{weller2025promptriever}.

Existing instruction-following retrieval benchmarks show that recent models can shift rankings in response to changed instructions \citep{weller2025followir,oh2024instructir,zhou2025beyond, sun2024mair,song2025ifir}. \citet{maheshwari2026exploration} find that ranking gains in aspect-conditioned exploration do not necessarily track instruction sensitivity. SwitchIR isolates the corresponding decision failure under fixed queries and candidate pairs, with the target changed solely through relational role binding.

\paragraph{Diagnostic and behavioral evaluation.}
Our work is also related to targeted diagnostic evaluation in NLP and LLMs. Behavioral testing frameworks show that aggregate held-out scores can miss systematic failures on specific capabilities \citep{ribeiro2020beyond}. Other diagnostics expose shallow heuristics and sensitivity to controlled structural perturbations, including lexical-overlap heuristics in NLI and directional failures in relational knowledge \citep{mccoy2019right,berglund2024reversal}. Knowledge graph methods such as TransE and RotatE explicitly represent asymmetric relations in a symbolic triple space \citep{bordes2013translating,sun2019rotate}. Dense bi-encoders can encode direction through instruction-conditioned query and document representations; successful reversal additionally requires the induced query change to align strongly with the document contrast. SwitchIR measures this requirement at the decision level by asking whether a model moves the ranking boundary between two topically plausible documents when only relational structure changes.

\section{SwitchIR: A Relational Diagnostic}

\subsection{Motivation}

SwitchIR isolates instruction-conditioned pairwise preference under a fixed query and two mutually hard-negative documents. The instruction determines which document should rank higher, while the topic and candidate pair remain unchanged. This minimal setting removes corpus-level document recovery and candidate-composition effects, and tests whether the retriever can distinguish the requested relation direction once both plausible candidates are available. Both candidates are exposed under every instruction condition, so category differences reflect preference resolution rather than candidate availability. A failure therefore indicates that the model cannot reliably rank the requested direction above its reversed counterpart. The three categories distinguish whether the preferred document can be identified from visible attributes, explicit operator--feature binding, or directed relational structure.

\textbf{Constraint Shift} changes the positive constraint that the preferred document must satisfy, such as audience, aspect, region, evidence type, or use case. The two documents have visibly different attribute content, making this category the lexical/attribute-cue positive control.

\textbf{Explicit Filtering} is a bag-of-words-matched control. The two opposite-target instructions preserve the same content and feature-word inventory while exchanging which feature is included and which is excluded; the document pair makes the same opposing role assignment. The model must bind include/exclude to the corresponding feature. Each target is expressed in include-first and exclude-first order, providing an order-controlled realization with unchanged truth conditions.

\textbf{Relational Flip}, the key diagnostic focus of SwitchIR, fixes the query, entities, and mutually hard-negative document pair while reversing only the relation direction or semantic-role assignment requested by the instruction. Each direction has a canonical and a meaning-preserving equivalent realization. Both candidates remain lexically plausible, so the retriever must match the directed relation in the instruction to the one expressed in the document. We instantiate Relational Flip across six relation types: \textbf{Causal Influence} distinguishes cause from effect; \textbf{Temporal Order} distinguishes the earlier event from the later event; \textbf{Scalar Comparison} distinguishes larger from smaller, or more from less; \textbf{Containment/Hierarchy} distinguishes a container from its contained item, or a higher from a lower role; \textbf{Directed Action} distinguishes the agent of an action from its patient; and \textbf{Transfer/Information Flow} distinguishes giver from receiver, or source from destination.

Together, the three categories separate success based on visible attributes, success under matched lexical content but reversed include/exclude roles, and success under directed relational structure.

\subsection{Task Definition}

Let \(R_\theta\) be a retrieval model with parameters \(\theta\). Given a topic query \(q\), an instruction \(i\), and a candidate document \(d\), it assigns a real-valued relevance score \(s_\theta(q,i,d)\). This definition covers lexical models, embedding retrievers, late-interaction retrievers, and rerankers; Section~\ref{sec:paired-ranking-geometry} introduces the normalized embedding form used for the representation analysis.

The basic unit of SwitchIR is an instruction-contrast group \(g=(q_g,d_{g,a},d_{g,b},\mathcal{I}_g^a,\mathcal{I}_g^b)\), where \(q_g\) is the shared query, \(d_{g,a}\) and \(d_{g,b}\) form a minimal document pair, and the two instruction sets are defined according to their gold preferences: \(d_{g,a}\succ_{q_g,i}d_{g,b}\) for every \(i\in\mathcal{I}_g^a\), and \(d_{g,b}\succ_{q_g,i}d_{g,a}\) for every \(i\in\mathcal{I}_g^b\). The complete set of instruction conditions for group \(g\) is \(\mathcal{I}_g=\mathcal{I}_g^a\cup\mathcal{I}_g^b\).

Each preference set contains one designated anchor instruction, denoted by \(i_g^a\in\mathcal{I}_g^a\) and \(i_g^b\in\mathcal{I}_g^b\).
For any \(i\in\mathcal{I}_g\), the preferred and non-preferred documents are therefore
\[
\bigl(d_g^+(i),d_g^-(i)\bigr)
=
\begin{cases}
\bigl(d_{g,a},d_{g,b}\bigr),
& i\in\mathcal{I}_g^a,\\[2mm]
\bigl(d_{g,b},d_{g,a}\bigr),
& i\in\mathcal{I}_g^b.
\end{cases}
\]

Constraint Shift contains one instruction in each preference set. Explicit Filtering contains two instructions in each preference set: the include-first condition is used as the anchor, and the corresponding exclude-first condition is an additional realization of the same preference. Relational Flip likewise contains two instructions in each preference set: the canonical condition is used as the anchor, and its meaning-preserving equivalent is an additional realization. All conditions belonging to the same group are assigned to the same data split to prevent condition-level leakage.

\subsection{Data Construction}

SwitchIR contains two complementary variants: a controlled set for diagnostic isolation and a realistic set with more natural surface realization. Separate reporting preserves the distinction between them.

Dataset statistics and detailed controlled and realistic examples appear in Appendix~\ref{sec:switchir-data-appendix}.

For the controlled set, templates hold vocabulary and structure as fixed as possible. The two documents in each group share the topic and central entities but differ in the attribute, include/exclude role, or directed relation that determines the gold preference.

For the realistic set, we select topics that admit natural scenarios and use LLM-assisted rewrites to express the queries and documents as natural information needs and passages while preserving the contrast and gold labels. The authors manually reviewed every resulting group for instruction--document consistency and label correctness.

\subsection{Evaluation Metrics}

For each group \(g\) and instruction condition \(i\in\mathcal{I}_g\), define the condition-level correctness indicator
\[
c_g(i)
=
\mathbf{1}
\left[
s_\theta\bigl(q_g,i,d_g^+(i)\bigr)
>
s_\theta\bigl(q_g,i,d_g^-(i)\bigr)
\right].
\]
The strict inequality ensures that a tied score is treated as a failure to resolve the intended preference.

The correctness indicators for the two anchor conditions are \(c_g^{\mathrm{anc},a}=c_g(i_g^a)\) and \(c_g^{\mathrm{anc},b}=c_g(i_g^b)\). To additionally require correctness across all realizations of each preference, define \(c_g^{\mathrm{all},a}=\prod_{i\in\mathcal{I}_g^a}c_g(i)\) and \(c_g^{\mathrm{all},b}=\prod_{i\in\mathcal{I}_g^b}c_g(i)\).
Thus, \(c_g^{\mathrm{all},a}=1\) only when the retriever ranks \(d_{g,a}\) above \(d_{g,b}\) under every instruction in \(\mathcal{I}_g^a\), and \(c_g^{\mathrm{all},b}=1\) only when it ranks \(d_{g,b}\) above \(d_{g,a}\) under every instruction in \(\mathcal{I}_g^b\).

Let \(G\) denote the set of groups in the category and dataset variant being evaluated. We compute four metrics:
\begin{align}
\operatorname{PairAcc}
&=
\frac{1}{2|G|}
\sum_{g\in G}
\left(
c_g^{\mathrm{anc},a}
+
c_g^{\mathrm{anc},b}
\right),
\\
\operatorname{FlipBoth}
&=
\frac{1}{|G|}
\sum_{g\in G}
c_g^{\mathrm{anc},a}
c_g^{\mathrm{anc},b},
\\
\operatorname{ParaAll}
&=
\frac{1}{2|G|}
\sum_{g\in G}
\left(
c_g^{\mathrm{all},a}
+
c_g^{\mathrm{all},b}
\right),
\\
\operatorname{ContrastAll}
&=
\frac{1}{|G|}
\sum_{g\in G}
c_g^{\mathrm{all},a}
c_g^{\mathrm{all},b}.
\end{align}

PairAcc measures average accuracy over the two anchor preferences, whereas FlipBoth measures the proportion of groups for which both opposite anchor preferences are correctly resolved. ParaAll and ContrastAll impose the corresponding requirements over all realizations of each preference. All four metrics weight groups equally, preventing categories with more instruction realizations from receiving greater influence.

For Constraint Shift, both \(\mathcal{I}_g^a\) and \(\mathcal{I}_g^b\) are singleton sets. Consequently, \(c_g^{\mathrm{all},a}=c_g^{\mathrm{anc},a}\) and \(c_g^{\mathrm{all},b}=c_g^{\mathrm{anc},b}\), so ParaAll and ContrastAll reduce to PairAcc and FlipBoth, respectively.

\section{Experiments}

\begin{table*}[!t]
\caption{Zero-shot results on the controlled set. PAcc, FBoth, PAll, and CAll abbreviate PairAcc, FlipBoth, ParaAll, and ContrastAll. Random values assume independent uniform selection; rerankers form a separate block.}
\centering

\setlength{\tabcolsep}{1.2pt}
\begin{tabular}{@{}llcccccccccc@{}}
\toprule
Type & Model & \multicolumn{2}{c}{Constraint Shift} & \multicolumn{4}{c}{Explicit Filtering} & \multicolumn{4}{c}{Relational Flip} \\
\cmidrule(lr){3-4}\cmidrule(lr){5-8}\cmidrule(lr){9-12}
& & PAcc & FBoth & PAcc & FBoth & PAll & CAll & PAcc & FBoth & PAll & CAll \\
\midrule
Random & Random
& \scorecell{50}{0.500} & \scorecell{25}{0.250}
& \scorecell{50}{0.500} & \scorecell{25}{0.250} & \scorecell{25}{0.250} & \scorecell{6}{0.063}
& \scorecell{50}{0.500} & \scorecell{25}{0.250} & \scorecell{25}{0.250} & \scorecell{6}{0.063} \\
\midrule
Lexical & BM25
& \scorecell{100}{1.000} & \scorecell{100}{1.000}
& \scorecell{0}{0.000} & \scorecell{0}{0.000} & \scorecell{0}{0.000} & \scorecell{0}{0.000}
& \scorecell{0}{0.000} & \scorecell{0}{0.000} & \scorecell{0}{0.000} & \scorecell{0}{0.000} \\
\midrule
Retriever & Qwen3-Emb-0.6B & \scorecell{99}{0.990} & \scorecell{98}{0.980} & \scorecell{94}{0.945} & \scorecell{89}{0.890} & \scorecell{90}{0.900} & \scorecell{80}{0.800} & \scorecell{49}{0.493} & \scorecell{21}{0.210} & \scorecell{28}{0.279} & \scorecell{0}{0.003} \\
& Qwen3-Emb-4B & \scorecell{97}{0.970} & \scorecell{94}{0.940} & \scorecell{99}{0.990} & \scorecell{98}{0.980} & \scorecell{99}{0.990} & \scorecell{98}{0.980} & \scorecell{52}{0.516} & \scorecell{14}{0.142} & \scorecell{39}{0.391} & \scorecell{0}{0.000} \\
& Qwen3-Emb-8B & \scorecell{100}{1.000} & \scorecell{100}{1.000} & \scorecell{98}{0.975} & \scorecell{95}{0.950} & \scorecell{97}{0.970} & \scorecell{94}{0.940} & \scorecell{56}{0.561} & \scorecell{26}{0.255} & \scorecell{38}{0.376} & \scorecell{2}{0.020} \\
& E5-Mistral & \scorecell{100}{1.000} & \scorecell{100}{1.000} & \scorecell{98}{0.985} & \scorecell{97}{0.970} & \scorecell{97}{0.970} & \scorecell{94}{0.940} & \scorecell{57}{0.572} & \scorecell{23}{0.229} & \scorecell{44}{0.437} & \scorecell{4}{0.045} \\
& GritLM & \scorecell{100}{1.000} & \scorecell{100}{1.000} & \scorecell{98}{0.985} & \scorecell{97}{0.970} & \scorecell{98}{0.980} & \scorecell{96}{0.960} & \scorecell{59}{0.586} & \scorecell{24}{0.245} & \scorecell{44}{0.442} & \scorecell{3}{0.032} \\
& Promptriever & \scorecell{100}{1.000} & \scorecell{100}{1.000} & \scorecell{100}{1.000} & \scorecell{100}{1.000} & \scorecell{100}{1.000} & \scorecell{100}{1.000} & \scorecell{52}{0.525} & \scorecell{15}{0.147} & \scorecell{40}{0.403} & \scorecell{0}{0.000} \\
\midrule
Late interaction & BGE-M3-ColBERT & \scorecell{99}{0.990} & \scorecell{98}{0.980} & \scorecell{94}{0.940} & \scorecell{88}{0.880} & \scorecell{92}{0.920} & \scorecell{84}{0.840} & \scorecell{53}{0.532} & \scorecell{27}{0.268} & \scorecell{30}{0.298} & \scorecell{0}{0.003} \\
\midrule
Reranker & Qwen3-Rerank-0.6B & \scorecell{100}{1.000} & \scorecell{100}{1.000} & \scorecell{100}{1.000} & \scorecell{100}{1.000} & \scorecell{100}{1.000} & \scorecell{100}{1.000} & \scorecell{66}{0.658} & \scorecell{46}{0.462} & \scorecell{41}{0.408} & \scorecell{10}{0.105} \\
& Qwen3-Rerank-4B & \scorecell{100}{1.000} & \scorecell{100}{1.000} & \scorecell{100}{1.000} & \scorecell{100}{1.000} & \scorecell{100}{1.000} & \scorecell{100}{1.000} & \scorecell{80}{0.805} & \scorecell{62}{0.623} & \scorecell{63}{0.632} & \scorecell{29}{0.292} \\
\bottomrule
\end{tabular}
\label{tab:main-results-controlled}
\end{table*}

\begin{table}[!t]
\caption{Primary joint metrics on the realistic set. FBoth is used for Constraint Shift and CAll for Explicit Filtering and Relational Flip. Full results appear in Appendix Table~\ref{tab:full-results-realistic}.}
\centering
\setlength{\tabcolsep}{3pt}
\begin{tabular}{@{}lccc@{}}
\toprule
Model & C.\ Shift & E.\ Filter & R.\ Flip \\
& FBoth & CAll & CAll \\
\midrule
Random & \scorecell{25}{0.250} & \scorecell{6}{0.063} & \scorecell{6}{0.063} \\
\midrule
BM25 & \scorecell{0}{0.000} & \scorecell{0}{0.000} & \scorecell{0}{0.000} \\
\midrule
Qwen3-Emb-0.6B & \scorecell{85}{0.850} & \scorecell{65}{0.650} & \scorecell{2}{0.020} \\
Qwen3-Emb-4B & \scorecell{100}{1.000} & \scorecell{85}{0.850} & \scorecell{10}{0.100} \\
Qwen3-Emb-8B & \scorecell{100}{1.000} & \scorecell{75}{0.750} & \scorecell{12}{0.120} \\
E5-Mistral & \scorecell{95}{0.950} & \scorecell{45}{0.450} & \scorecell{12}{0.120} \\
GritLM & \scorecell{100}{1.000} & \scorecell{90}{0.900} & \scorecell{18}{0.180} \\
Promptriever & \scorecell{95}{0.950} & \scorecell{100}{1.000} & \scorecell{1}{0.010} \\
\midrule
BGE-M3-ColBERT & \scorecell{85}{0.850} & \scorecell{15}{0.150} & \scorecell{4}{0.040} \\
\midrule
Qwen3-Rerank-0.6B & \scorecell{65}{0.650} & \scorecell{100}{1.000} & \scorecell{9}{0.090} \\
Qwen3-Rerank-4B & \scorecell{100}{1.000} & \scorecell{100}{1.000} & \scorecell{47}{0.470} \\
\bottomrule
\end{tabular}
\label{tab:main-results-realistic}
\end{table}

\subsection{Experimental Setup}

We evaluate all models in a zero-shot setting. For each SwitchIR group, the model receives a query, an instruction, and the two candidate documents, and must select the document preferred under that instruction. Because the candidate pair is fixed within each group, this setup evaluates instruction-conditioned pairwise preference rather than corpus-level recall.

\paragraph{Scoring.}
The embedding retrievers use cosine similarity between the instruction-conditioned query embedding and each document embedding. BGE-M3-ColBERT uses token-level late-interaction scores, BM25 uses lexical relevance scores, and the Qwen3 rerankers assign cross-encoder relevance scores to each instruction-conditioned query--document pair. In every case, the candidate with the higher score is predicted, and a score tie counts as an error.

\paragraph{Models.}
We evaluate embedding retrievers and instruction-aware embedding models, including Qwen3-Embedding variants, E5-Mistral, GritLM, and Promptriever \citep{wang2022text,wang2024improving,muennighoff2025generative,zhang2025qwen3,weller2025promptriever}. Tables and figures abbreviate Qwen3-Embedding as Qwen3-Emb. BGE-M3-ColBERT is evaluated as a late-interaction retriever. We include BM25 as a lexical baseline \citep{robertson2009probabilistic} and Qwen3 rerankers as non-bi-encoder references \citep{nogueira2019passage,zhang2025qwen3}. Rerankers are reported separately from the retrieval models. For every instruction-aware neural model, we follow its original model-specific instruction template. Appendix~\ref{sec:model-instruction-templates} records the templates used for each model.

\paragraph{Datasets and metrics.}
We report the controlled set (700 groups and 2,700 instruction conditions) and realistic set (140 groups and 520 conditions) separately, without a pooled score. Constraint Shift uses PairAcc and FlipBoth. Explicit Filtering and Relational Flip additionally use ParaAll and ContrastAll, with ParaAll denoting consistency across the two realizations of the same preference. Under independent random choice, PairAcc and FlipBoth have expectations \(0.5\) and \(0.25\); for the four-condition categories, ParaAll and ContrastAll have expectations \(0.25\) and \(0.0625\). For compact realistic-set reporting, Table~\ref{tab:main-results-realistic} uses FlipBoth for Constraint Shift and ContrastAll for Explicit Filtering and Relational Flip as the category-specific primary joint metrics; Appendix Table~\ref{tab:full-results-realistic} reports the complete metric set. No training runs are included in this zero-shot comparison.

\subsection{Main SwitchIR Results}

Tables~\ref{tab:main-results-controlled} and~\ref{tab:main-results-realistic} report the controlled and realistic results. On the controlled benchmark, performance decreases sharply as the preferred document must be identified from less explicit cues. Averaged over the seven non-reranker neural retrieval runs, Constraint Shift FlipBoth reaches \(0.986\), Explicit Filtering ContrastAll reaches \(0.923\), and Relational Flip ContrastAll reaches only \(0.015\). Constraint Shift can be resolved from visible attribute differences, while Explicit Filtering requires binding include/exclude operators to their corresponding features under a matched word inventory. The further collapse on Relational Flip shows that the difficulty cannot be explained by bag-of-words matching alone: models that successfully resolve explicit operator--feature roles still fail to reliably match general relational roles across instructions and documents. 


The realistic variant reproduces the same category ordering under more natural queries and passages. The corresponding average scores are \(0.943\) for Constraint Shift, \(0.679\) for Explicit Filtering, and \(0.084\) for Relational Flip. Relational Flip therefore remains substantially more difficult after removing the highly controlled template expressions, showing that the observed gap is not specific to the surface form of the controlled data.

The failure also persists across retrieval architectures. BM25 performs well on controlled Constraint Shift but obtains no joint success on the matched-word categories, confirming that lexical matching alone cannot resolve either role-binding task. Neural embedding and late-interaction retrievers substantially improve over lexical retrieval on individual decisions, yet remain unreliable on joint Relational Flip. Cross-encoder reranking provides stronger results, especially at larger scale, but does not eliminate the gap: Qwen3-Rerank-4B reaches Relational Flip ContrastAll of \(0.292\) controlled and \(0.470\) realistic. Even this strongest reranker retains a \(0.53\) realistic ContrastAll gap between Explicit Filtering and Relational Flip. The difficulty therefore cannot be resolved through a simple change from dense embeddings to late interaction or reranking.

\subsection{Relation to FollowIR}
\label{sec:followir-gap}

FollowIR p-MRR measures full-corpus rank changes on its conventional \(100\times\) scale, whereas SwitchIR ContrastAll is a conjunction accuracy on fixed two-document groups. We therefore compare model ordering and diagnostic coverage.

Table~\ref{tab:followir-gap} uses FollowIR metrics and controlled separate instruction results. Promptriever has the highest FollowIR p-MRR but zero Relational Flip ContrastAll. In contrast, E5-Mistral and GritLM have near-zero FollowIR p-MRR but higher Relational Flip consistency. The divergent model ordering shows that the two diagnostics capture distinct aspects of instruction following.

\begin{table}[t]
\centering
\small
\caption{FollowIR p-MRR versus controlled Relational Flip ContrastAll.}
\label{tab:followir-gap}
\setlength{\tabcolsep}{3pt}
\begin{tabular}{lcc}
\toprule
Model & FollowIR & Rel. Flip \\
& p-MRR & ContrastAll \\
\midrule
Qwen3-Emb-0.6B & 2.6 & 0.003 \\
Qwen3-Emb-4B   & 10.3 & 0.000 \\
Qwen3-Emb-8B   & 7.9 & 0.020 \\
E5-Mistral     & 0.0 & 0.045 \\
GritLM         & 0.7 & 0.032 \\
Promptriever   & 12.8 & 0.000 \\
\bottomrule
\end{tabular}
\end{table}

\section{Analysis}

We analyze Relational Flip through the paired ranking decisions evaluated by SwitchIR. An instruction-induced query change supports the target switch through its alignment with the document contrast, while a stable preference for one member of the pair can oppose that switch. We therefore decompose each retrieval margin into a direction-invariant pair preference and a direction-sensitive query--document interaction. The decomposition links every failed flip to a measurable imbalance between instruction response and pair preference. This geometry applies specifically to Relational Flip, whose fixed query and document pair exchange gold documents under opposite directions. We compute it separately for every group and realization, without averaging embeddings across groups.

\subsection{Paired Ranking Geometry}
\label{sec:paired-ranking-geometry}

Let \(t\in\{c,e\}\) index the canonical and equivalent realizations of a Relational Flip group \(g\). Let \(i_{AB}^{(t)}\) and \(i_{BA}^{(t)}\) denote the instructions requesting directions \(A\!\rightarrow\!B\) and \(B\!\rightarrow\!A\), and let \(d_{AB}=d_g^+(i_{AB}^{(t)})\) and \(d_{BA}=d_g^+(i_{BA}^{(t)})\) denote their gold documents; the two realizations share these documents. For the four embedding retrievers analyzed here, let \(h_\theta(q,i)\) and \(h_\theta(d)\) denote the instruction-conditioned query embedding and document embedding. Using each model's original instruction template, define the normalized query embeddings as
\begin{align}
x_{AB}^{(t)}
&=
\frac{h_\theta(q_g,i_{AB}^{(t)})}
{\lVert h_\theta(q_g,i_{AB}^{(t)})\rVert_2},
\\
x_{BA}^{(t)}
&=
\frac{h_\theta(q_g,i_{BA}^{(t)})}
{\lVert h_\theta(q_g,i_{BA}^{(t)})\rVert_2}.
\end{align}
The corresponding normalized document embeddings are
\[
y_{AB}
=
\frac{h_\theta(d_{AB})}{\lVert h_\theta(d_{AB})\rVert_2},
\qquad
y_{BA}
=
\frac{h_\theta(d_{BA})}{\lVert h_\theta(d_{BA})\rVert_2}.
\]
The subscripts identify the requested relation direction and its gold document, not a batch position. All four vectors have unit norm, so cosine scoring reduces to \(s(x,y)=x^\top y\), with \(\lVert x\rVert_2=\lVert y\rVert_2=1\).

For a fixed realization \(t\), the two gold-document margins are
\begin{align}
m_{AB}^{(t)}
&=
s(x_{AB}^{(t)},y_{AB})
-
s(x_{AB}^{(t)},y_{BA}),
\\
m_{BA}^{(t)}
&=
s(x_{BA}^{(t)},y_{BA})
-
s(x_{BA}^{(t)},y_{AB}).
\end{align}
A direction is correct exactly when its margin is strictly positive; a tied score remains an error under the SwitchIR evaluation rule.

Suppressing \(t\) for readability, define \(\mu_x=(x_{AB}+x_{BA})/2\), \(\Delta_x=x_{AB}-x_{BA}\), and \(\Delta_y=y_{AB}-y_{BA}\). We then set \(b=\mu_x^\top\Delta_y\) and \(r=\frac{1}{2}\Delta_x^\top\Delta_y\).
The term \(b\) is the \emph{direction-invariant pair preference}: it measures whether the average query favors \(y_{AB}\) or \(y_{BA}\). Its joint dependence on \(\mu_x\) and \(\Delta_y\) makes it a query--document-pair statistic. The term \(r\) is the \emph{direction-sensitive query--document interaction signal}. Positive \(r\) indicates that the query change is aligned with the required target switch, while negative \(r\) indicates an anti-aligned change.

Since \(x_{AB}=\mu_x+\frac{1}{2}\Delta_x\) and \(x_{BA}=\mu_x-\frac{1}{2}\Delta_x\), the margins decompose exactly as \(m_{AB}=b+r\) and \(m_{BA}=-b+r\). Both directions are therefore correct if and only if \(m_{AB}>0\land m_{BA}>0\), equivalently \(r>|b|\). The decision surplus \(d=r-|b|=\min(m_{AB},m_{BA})\) is thus positive exactly when that realization switches correctly in both directions. A zero surplus includes at least one tie and is a failure. This equivalence is pointwise: we first evaluate it for each group--realization observation and then aggregate the resulting indicators. Comparisons between mean \(r\) and mean \(|b|\) are descriptive summaries and do not replace the condition-level success rate.

The decomposition yields four diagnostic regions. Sufficient aligned signal satisfies \(r>|b|\); aligned but insufficient signal satisfies \(0<r\le|b|\); anti-aligned signal satisfies \(r<0\); and \(r=0\) provides no interaction contribution for switching the target.

\subsection{Factors Governing the Interaction Signal}

The interaction term factorizes as
\begin{align}
r
&=
\frac{1}{2}
\lVert\Delta_x\rVert_2
\lVert\Delta_y\rVert_2
\cos\theta,
\\
\cos\theta
&=
\frac{\Delta_x^\top\Delta_y}
{\lVert\Delta_x\rVert_2\lVert\Delta_y\rVert_2}.
\end{align}
Here, \(\lVert\Delta_x\rVert_2\) is the query direction contrast, \(\lVert\Delta_y\rVert_2\) is the document contrast, and \(\cos\theta\) is their cross-space alignment. Unit normalization bounds both contrast norms to \([0,2]\), with \(\lVert\Delta_x\rVert_2^2=2\bigl(1-\cos(x_{AB},x_{BA})\bigr)\).
This factorization gives a direct attribution. Query contrast and document contrast scale the interaction \(r\), while cross-space alignment determines whether that interaction supports or opposes the required reversal. If either contrast vector is zero, the inner-product definition gives \(r=0\); if their alignment is non-positive, the interaction cannot satisfy \(r>|b|\). Switching depends on the joint interaction relative to the direction-invariant pair preference: even a large query shift, well-separated documents, or positive alignment can leave \(r\leq |b|\).

We additionally measure whether the two meaning-equivalent realizations induce consistent direction changes with \(\rho_{\mathrm{real}}=\cos\!\bigl(\Delta_x^{(c)},\Delta_x^{(e)}\bigr)\).
This is a representation-level realization-stability measure. It is distinct from ParaAll, which is computed from the four actual ranking decisions. Together, these factors distinguish weak query response, weak document contrast, cross-space misalignment, and an overly strong direction-invariant pair preference.

\subsection{Empirical Geometry}

\begin{figure}[!t]
\centering
\includegraphics[width=\linewidth]{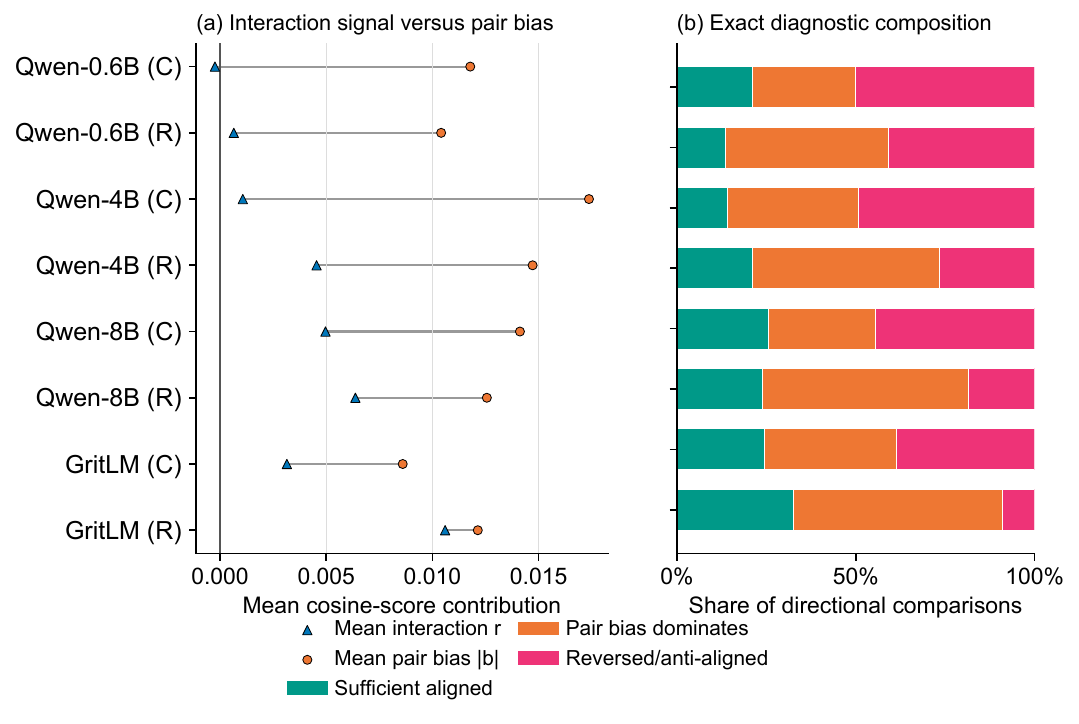}
\caption{Exact Relational Flip geometry for the four embedding retrievers. C and R denote controlled and realistic data. Panel (a) compares mean interaction signal \(r\) with mean pair-preference magnitude \(|b|\). Panel (b) partitions group--realization observations using the pointwise regions defined above. Each model contributes 1,200 controlled observations and 200 realistic observations.}
\label{fig:geometry-signal-vs-bias}
\end{figure}

Figure~\ref{fig:geometry-signal-vs-bias} summarizes the geometry for Qwen3-Emb-0.6B, Qwen3-Emb-4B, Qwen3-Emb-8B, and GritLM. This analysis covers these four embedding retrievers only. Each observation is one Relational Flip group under one realization, yielding 1,200 controlled and 200 realistic observations per model.

Mean \(|b|\) exceeds mean \(r\) in every model--dataset cell. More importantly, the pointwise partition shows that sufficient aligned signal occurs for \(14.2\%\)--\(25.5\%\) of controlled observations and \(13.5\%\)--\(32.5\%\) of realistic observations. Aligned but insufficient interaction is more common in every cell, accounting for \(28.8\%\)--\(36.9\%\) of controlled observations and \(45.5\%\)--\(58.5\%\) of realistic observations. Pair-preference dominance is therefore the principal zero-shot failure mode for these four retrievers: direction-sensitive changes are common, but their aligned interaction often fails to exceed \(|b|\). Query contrast, document contrast, and alignment contribute jointly through \(r\).

\subsection{Diagnostic Training}
\label{sec:diagnostic-training}

We use direction-sensitive training as a diagnostic intervention. The experiments test whether the existing encoders can learn Relational Flip and whether the learned behavior extends to unseen relation families. Both experiments fine-tune Qwen3-Embedding-0.6B and Qwen3-Embedding-4B with LoRA using controlled training data only. For each training instance \(x\), the correct-direction document is \(y^+\), the reversed document is \(y^-\), and the pairwise objective is \(\mathcal{L}_{\mathrm{rank}}=-\log \sigma\bigl((s(x,y^+)-s(x,y^-))/\tau\bigr)\), where \(\sigma\) is the logistic function. Training covers both query realizations, both document realizations, and both relation directions. Realistic data are reserved for external evaluation. We report ContrastAll in percent, first macro-averaged equally over relation families within each seed and then summarized as mean \(\pm\) sample standard deviation over seeds 42--46. Full optimization and selection details appear in Appendix~\ref{sec:training-details}.

\paragraph{Single-relation training.}
Each run uses the 70 controlled training groups from one of the six relation families. Checkpoints are selected on controlled development data from that family. Same-relation evaluation averages the six matched train--test conditions, whereas cross-relation evaluation averages the 30 directed unmatched conditions.

\begin{figure}[!t]
\centering
\includegraphics[width=\linewidth]{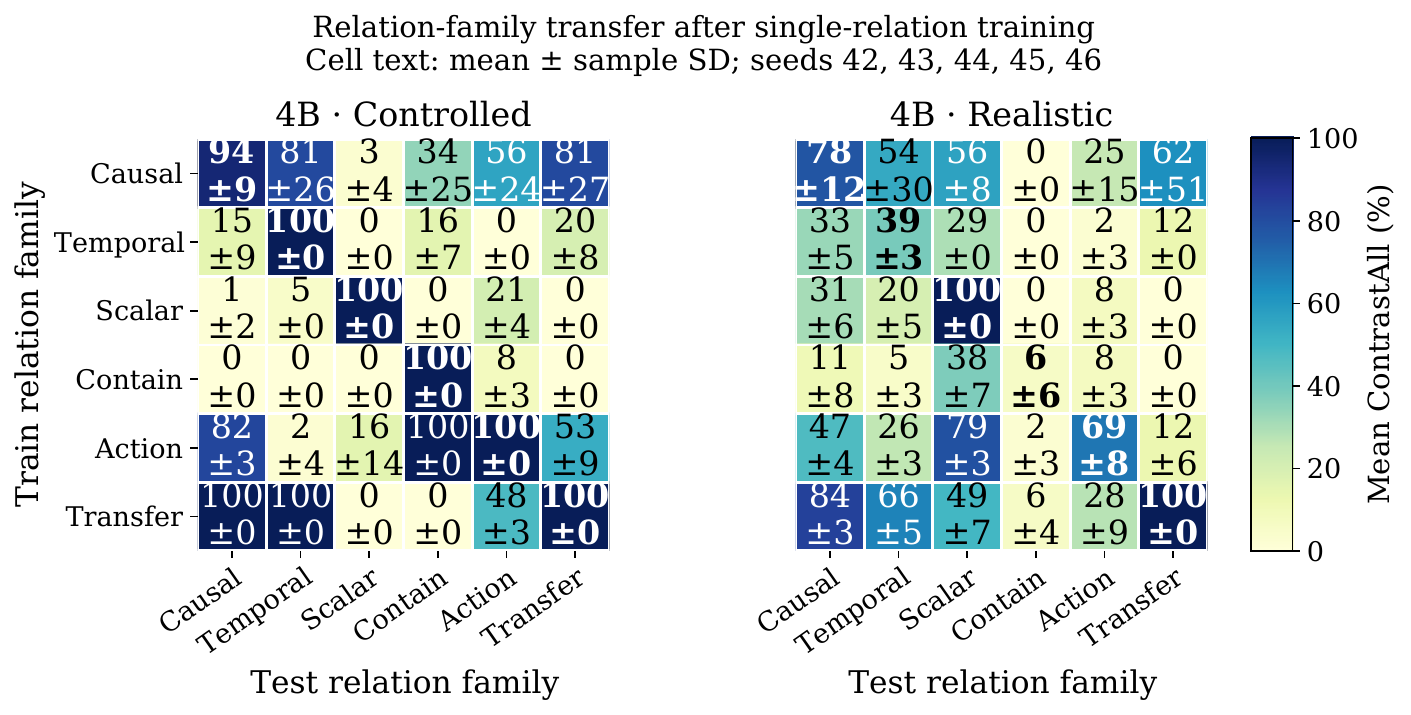}
\caption{Single-relation Qwen3-Embedding-4B transfer. Rows/columns are training/test families; cells show mean ContrastAll (\%) \(\pm\) sample SD over five seeds.}
\label{fig:single-relation-transfer}
\end{figure}

Figure~\ref{fig:single-relation-transfer} shows that one-family supervision learns the controlled training family (\(99.0\%\) diagonal average) but transfers unevenly to others (\(28.1\%\) off-diagonal average), with strong directional asymmetries. On realistic data, these averages fall to \(65.2\%\) and \(26.4\%\), and the diagonal ranges from \(6\%\) for Containment/Hierarchy to \(100\%\) for Scalar Comparison and Transfer/Information Flow. Single-family training is therefore relation specific and depends on direction, family, and surface realization.

\paragraph{Leave-one-relation-out training.}
We next hold out each relation family in turn and train on 350 controlled groups from the other five families. The held-out family's training and development groups are excluded from parameter updates and checkpoint selection. Evaluation is restricted to that unseen family and its realistic counterpart.

\begin{table}[!t]
\centering
\small
\setlength{\tabcolsep}{3pt}
\caption{Leave-one-relation-out diagnostic training. Values are ContrastAll percentages; held-out results are mean \(\pm\) sample standard deviation over five seeds.}
\label{tab:leave-one-out-training}
\begin{tabular}{lcccc}
\toprule
& \multicolumn{2}{c}{Controlled} & \multicolumn{2}{c}{Realistic} \\
\cmidrule(lr){2-3}\cmidrule(lr){4-5}
Size & Frozen & Held out & Frozen & Held out \\
\midrule
0.6B & 0.0 & $35.7 \mathbin{\pm} 5.2$ & 2.0 & $13.5 \mathbin{\pm} 3.3$ \\
4B & 0.0 & $92.5 \mathbin{\pm} 6.6$ & 9.9 & $70.1 \mathbin{\pm} 3.9$ \\
\bottomrule
\end{tabular}
\end{table}

The 4B model transfers the learned directional pattern to unseen relations, reaching \(92.5\%\) controlled and \(70.1\%\) realistic ContrastAll in Table~\ref{tab:leave-one-out-training}. The corresponding 0.6B results are \(35.7\%\) and \(13.5\%\). The larger model therefore achieves strong held-out generalization under this protocol, with lower and more heterogeneous transfer on realistic data. Its realistic scores range from \(17.6\%\) for Containment/Hierarchy to \(97.5\%\) for Directed Action (Figure~\ref{fig:training-heterogeneity}).

Together, the experiments establish learnability, relation specificity under single-family supervision, and a pronounced model-size difference in multi-relation generalization.

\section{Conclusion}

SwitchIR reveals a sharp divide in instruction-following retrieval. Retrievers reliably switch preferences under explicit filtering and attribute constraints, yet struggle when a similarly small change reverses the same entities' relational roles. Success on cue-rich tasks therefore does not imply reliable relational role binding. The models often encode the requested direction change, but the resulting document interaction is too weak to overturn their existing pair preference. Relational retrieval thus depends on both a direction-sensitive signal and sufficient interaction strength to cross the decision boundary. Diagnostic training confirms that relational switching is learnable: single-relation supervision remains relation-specific, whereas multi-relation supervision transfers much more strongly in the higher-capacity encoder.

\section{Limitations}

SwitchIR focuses on directional relational structure, where the same entities are assigned to reversed semantic roles. This controlled setting does not cover other forms of structurally specified relevance, such as nested constraints, multi-hop relations, or interactions among multiple conditions.

\bibliography{references}

@book{robertson2009probabilistic,
  title={The probabilistic relevance framework: BM25 and beyond},
  author={Robertson, Stephen and Zaragoza, Hugo},
  volume={4},
  year={2009},
  publisher={Now Publishers Inc}
}

@article{nogueira2019passage,
  title={Passage Re-ranking with BERT},
  author={Nogueira, Rodrigo and Cho, Kyunghyun},
  journal={arXiv preprint arXiv:1901.04085},
  year={2019}
}

@inproceedings{karpukhin2020dense,
  title={Dense passage retrieval for open-domain question answering},
  author={Karpukhin, Vladimir and Oguz, Barlas and Min, Sewon and Lewis, Patrick and Wu, Ledell and Edunov, Sergey and Chen, Danqi and Yih, Wen-tau},
  booktitle={Proceedings of the 2020 conference on empirical methods in natural language processing (EMNLP)},
  pages={6769--6781},
  year={2020}
}

@article{xiong2020approximate,
  title={Approximate nearest neighbor negative contrastive learning for dense text retrieval},
  author={Xiong, Lee and Xiong, Chenyan and Li, Ye and Tang, Kwok-Fung and Liu, Jialin and Bennett, Paul and Ahmed, Junaid and Overwijk, Arnold},
  journal={arXiv preprint arXiv:2007.00808},
  year={2020}
}

@inproceedings{khattab2020colbert,
  title={Colbert: Efficient and effective passage search via contextualized late interaction over bert},
  author={Khattab, Omar and Zaharia, Matei},
  booktitle={Proceedings of the 43rd International ACM SIGIR conference on research and development in Information Retrieval},
  pages={39--48},
  year={2020}
}

@article{izacard2021unsupervised,
  title={Unsupervised dense information retrieval with contrastive learning},
  author={Izacard, Gautier and Caron, Mathilde and Hosseini, Lucas and Riedel, Sebastian and Bojanowski, Piotr and Joulin, Armand and Grave, Edouard},
  journal={arXiv preprint arXiv:2112.09118},
  year={2021}
}

@inproceedings{asai2023task,
  title={Task-aware retrieval with instructions},
  author={Asai, Akari and Schick, Timo and Lewis, Patrick and Chen, Xilun and Izacard, Gautier and Riedel, Sebastian and Hajishirzi, Hannaneh and Yih, Wen-tau},
  booktitle={Findings of the Association for Computational Linguistics: ACL 2023},
  pages={3650--3675},
  year={2023}
}

@inproceedings{pan2024i3,
  title={I3: I ntent-i ntrospective retrieval conditioned on i nstructions},
  author={Pan, Kaihang and Li, Juncheng and Wang, Wenjie and Fei, Hao and Song, Hongye and Ji, Wei and Lin, Jun and Liu, Xiaozhong and Chua, Tat-Seng and Tang, Siliang},
  booktitle={Proceedings of the 47th International ACM SIGIR Conference on Research and Development in Information Retrieval},
  pages={1839--1849},
  year={2024}
}

@inproceedings{su2023one,
  title={One embedder, any task: Instruction-finetuned text embeddings},
  author={Su, Hongjin and Shi, Weijia and Kasai, Jungo and Wang, Yizhong and Hu, Yushi and Ostendorf, Mari and Yih, Wen-tau and Smith, Noah A and Zettlemoyer, Luke and Yu, Tao},
  booktitle={Findings of the Association for Computational Linguistics: ACL 2023},
  pages={1102--1121},
  year={2023}
}

@article{wang2022text,
  title={Text embeddings by weakly-supervised contrastive pre-training},
  author={Wang, Liang and Yang, Nan and Huang, Xiaolong and Jiao, Binxing and Yang, Linjun and Jiang, Daxin and Majumder, Rangan and Wei, Furu},
  journal={arXiv preprint arXiv:2212.03533},
  year={2022}
}

@inproceedings{wang2024improving,
  title={Improving text embeddings with large language models},
  author={Wang, Liang and Yang, Nan and Huang, Xiaolong and Yang, Linjun and Majumder, Rangan and Wei, Furu},
  booktitle={Proceedings of the 62nd Annual Meeting of the Association for Computational Linguistics (Volume 1: Long Papers)},
  pages={11897--11916},
  year={2024}
}

@inproceedings{muennighoff2025generative,
  title={Generative representational instruction tuning},
  author={Muennighoff, Niklas and Su, Hongjin and Wang, Liang and Yang, Nan and Wei, Furu and Yu, Tao and Singh, Amanpreet and Kiela, Douwe},
  booktitle={International Conference on Learning Representations},
  volume={2025},
  pages={45544--45613},
  year={2025}
}

@article{zhang2025qwen3,
  title={Qwen3 embedding: Advancing text embedding and reranking through foundation models},
  author={Zhang, Yanzhao and Li, Mingxin and Long, Dingkun and Zhang, Xin and Lin, Huan and Yang, Baosong and Xie, Pengjun and Yang, An and Liu, Dayiheng and Lin, Junyang and others},
  journal={arXiv preprint arXiv:2506.05176},
  year={2025}
}

@article{oh2024instructir,
  title={Instructir: A benchmark for instruction following of information retrieval models},
  author={Oh, Hanseok and Lee, Hyunji and Ye, Seonghyeon and Shin, Haebin and Jang, Hansol and Jun, Changwook and Seo, Minjoon},
  journal={arXiv preprint arXiv:2402.14334},
  year={2024}
}

@inproceedings{weller2025followir,
  title={Followir: Evaluating and teaching information retrieval models to follow instructions},
  author={Weller, Orion and Chang, Benjamin and MacAvaney, Sean and Lo, Kyle and Cohan, Arman and Van Durme, Benjamin and Lawrie, Dawn and Soldaini, Luca},
  booktitle={Proceedings of the 2025 Conference of the Nations of the Americas Chapter of the Association for Computational Linguistics: Human Language Technologies (Volume 1: Long Papers)},
  pages={11926--11942},
  year={2025}
}

@inproceedings{weller2025promptriever,
  title={Promptriever: Instruction-trained retrievers can be prompted like language models},
  author={Weller, Orion and Van Durme, Ben and Lawrie, Dawn and Paranjape, Ashwin and Zhang, Yuhao and Hessel, Jack},
  booktitle={International Conference on Learning Representations},
  volume={2025},
  pages={17660--17683},
  year={2025}
}

@inproceedings{zhou2025beyond,
  title={Beyond content relevance: Evaluating instruction following in retrieval models},
  author={Zhou, Jianqun and Zheng, Yuanlei and Chen, Wei and Zheng, Qianqian and Zeyuan, Shang and Zhang, Wei and Meng, Rui and Shen, Xiaoyu},
  booktitle={International Conference on Learning Representations},
  volume={2025},
  pages={84965--84996},
  year={2025}
}

@inproceedings{sun2024mair,
  title={MAIR: A massive benchmark for evaluating instructed retrieval},
  author={Sun, Weiwei and Shi, Zhengliang and Long, Wu Jiu and Yan, Lingyong and Ma, Xinyu and Liu, Yiding and Cao, Min and Yin, Dawei and Ren, Zhaochun},
  booktitle={Proceedings of the 2024 Conference on Empirical Methods in Natural Language Processing},
  pages={14044--14067},
  year={2024}
}

@inproceedings{song2025ifir,
  title={IFIR: A comprehensive benchmark for evaluating instruction-following in expert-domain information retrieval},
  author={Song, Tingyu and Gan, Guo and Shang, Mingsheng and Zhao, Yilun},
  booktitle={Proceedings of the 2025 Conference of the Nations of the Americas Chapter of the Association for Computational Linguistics: Human Language Technologies (Volume 1: Long Papers)},
  pages={10186--10204},
  year={2025}
}

@inproceedings{ribeiro2020beyond,
  title={Beyond accuracy: Behavioral testing of NLP models with CheckList},
  author={Ribeiro, Marco Tulio and Wu, Tongshuang and Guestrin, Carlos and Singh, Sameer},
  booktitle={Proceedings of the 58th annual meeting of the association for computational linguistics},
  pages={4902--4912},
  year={2020}
}

@inproceedings{mccoy2019right,
  title={Right for the wrong reasons: Diagnosing syntactic heuristics in natural language inference},
  author={McCoy, R Thomas and Pavlick, Ellie and Linzen, Tal},
  booktitle={Proceedings of the 57th annual meeting of the association for computational linguistics},
  pages={3428--3448},
  year={2019}
}

@inproceedings{berglund2024reversal,
  title={The Reversal Curse: {LLM}s trained on ``{A} is {B}'' fail to learn ``{B} is {A}''},
  author={Berglund, Lukas and Tong, Meg and Kaufmann, Maximilian and Balesni, Mikita and Stickland, Asa and Korbak, Tomek and Evans, Owain},
  booktitle={International Conference on Learning Representations},
  volume={2024},
  pages={18623--18642},
  year={2024}
}

@article{zhuang2025towards,
  title={Towards Better Instruction Following Retrieval Models},
  author={Zhuang, Yuchen and Trinh, Aaron and Qiang, Rushi and Sun, Haotian and Zhang, Chao and Dai, Hanjun and Dai, Bo},
  journal={arXiv preprint arXiv:2505.21439},
  year={2025}
}

@article{zeng2026dual,
  title={Dual-View Training for Instruction-Following Information Retrieval},
  author={Zeng, Qingcheng and Yu, Puxuan and Mehta, Aman and Zhao, Fuheng and Samdani, Rajhans},
  journal={arXiv preprint arXiv:2604.18845},
  year={2026}
}

@inproceedings{bordes2013translating,
  title={Translating Embeddings for Modeling Multi-relational Data},
  author={Bordes, Antoine and Usunier, Nicolas and Garcia-Dur{\'a}n, Alberto and Weston, Jason and Yakhnenko, Oksana},
  booktitle={Advances in Neural Information Processing Systems},
  volume={26},
  year={2013}
}

@inproceedings{sun2019rotate,
  title={{RotatE}: Knowledge Graph Embedding by Relational Rotation in Complex Space},
  author={Sun, Zhiqing and Deng, Zhi-Hong and Nie, Jian-Yun and Tang, Jian},
  booktitle={International Conference on Learning Representations},
  year={2019}
}

@inproceedings{maheshwari2026exploration,
  title={Can Instructed Retrieval Models Really Support Exploration?},
  author={Maheshwari, Piyush and Mysore, Sheshera and Zamani, Hamed},
  booktitle={Proceedings of the 2026 ACM SIGIR Conference on Human Information Interaction and Retrieval},
  pages={508--513},
  year={2026},
  doi={10.1145/3786304.3787888}
}

\appendix
\section{SwitchIR Dataset Statistics and Examples}
\label{sec:switchir-data-appendix}

The three categories isolate different sources of instruction-conditioned preference changes. \textbf{Constraint Shift} changes a visible target attribute, such as audience or region. \textbf{Explicit Filtering} retains the feature vocabulary but exchanges which feature must be included and which must be excluded. \textbf{Relational Flip} retains the entities and relation vocabulary while reversing their semantic roles or directed relation. In the examples below, A and B mark opposite preferences; AB and BA mark opposite relation directions. Table~\ref{tab:switchir-example} gives one controlled group from each category, and Table~\ref{tab:switchir-realistic-example} gives a realistic Relational Flip group.

\begin{table*}[!tbp]
\centering
\small
\caption{Statistics of the controlled and realistic SwitchIR sets. A condition is one instruction--query pair with one gold document in its fixed two-document group.}
\setlength{\tabcolsep}{5pt}
\begin{tabular}{lrrrrrr}
\toprule
& \multicolumn{3}{c}{Controlled set} & \multicolumn{3}{c}{Realistic set} \\
\cmidrule(lr){2-4}\cmidrule(lr){5-7}
Category & Groups & Docs & Conditions & Groups & Docs & Conditions \\
\midrule
Constraint Shift & 50 & 100 & 100 & 20 & 40 & 40 \\
Explicit Filtering & 50 & 100 & 200 & 20 & 40 & 80 \\
Relational Flip & 600 & 1,200 & 2,400 & 100 & 200 & 400 \\
\midrule
Total & 700 & 1,400 & 2,700 & 140 & 280 & 520 \\
\bottomrule
\end{tabular}
\label{tab:switchir-stats}
\end{table*}

\begin{table*}[!t]
\centering
\caption{Controlled examples from the three SwitchIR categories, transcribed from the query, corpus, and relevance files. For compactness, the table shows the include-first conditions for Explicit Filtering and the canonical realizations for Relational Flip; the alternate conditions remain in the same groups.}
\footnotesize
\setlength{\tabcolsep}{3pt}
\begin{tabular}{p{0.12\textwidth}p{0.16\textwidth}p{0.35\textwidth}p{0.29\textwidth}}
\toprule
Category & Query & Instruction & Correct document \\
\midrule
Constraint Shift &
battery storage systems &
\textbf{A:} Retrieve documents about battery storage systems for teen users. &
\textbf{\(d_a\):} This passage discusses battery storage systems for teen users. Its analysis and examples consistently address teen users. \\
\cmidrule(lr){3-4}
&
&
\textbf{B:} Retrieve documents about battery storage systems for adult users. &
\textbf{\(d_b\):} This passage discusses battery storage systems for adult users. Its analysis and examples consistently address adult users. \\
\midrule
Explicit Filtering &
battery storage systems &
\textbf{A:} Retrieve documents about battery storage systems that include pricing information and exclude safety warnings. &
\textbf{\(d_a\):} This passage about battery storage systems includes pricing information and explicitly excludes safety warnings. \\
\cmidrule(lr){3-4}
&
&
\textbf{B:} Retrieve documents about battery storage systems that include safety warnings and exclude pricing information. &
\textbf{\(d_b\):} This passage about battery storage systems includes safety warnings and explicitly excludes pricing information. \\
\midrule
Relational Flip &
signal Aurora and response Kestrel in a municipal pilot &
\textbf{AB:} Find documents stating that signal Aurora causes response Kestrel in a municipal pilot. &
\textbf{\(d_a\):} In a municipal pilot, signal Aurora causes response Kestrel; the reverse direction is explicitly rejected. \\
\cmidrule(lr){3-4}
&
&
\textbf{BA:} Find documents stating that response Kestrel causes signal Aurora in a municipal pilot. &
\textbf{\(d_b\):} In a municipal pilot, response Kestrel causes signal Aurora; the reverse direction is explicitly rejected. \\
\bottomrule
\end{tabular}
\label{tab:switchir-example}
\end{table*}

\begin{table*}[!tbp]
\centering
\caption{A realistic Relational Flip example. The shared query and candidate pair remain fixed while the instruction reverses the requested causal direction.}
\footnotesize
\setlength{\tabcolsep}{4pt}
\begin{tabular}{p{0.16\textwidth}p{0.36\textwidth}p{0.40\textwidth}}
\toprule
Query & Instruction & Correct document \\
\midrule
What did the holiday review say about courier delays and refund requests? &
\textbf{AB:} Find the account stating that late courier handoffs caused customer refund requests in an online retailer's holiday postmortem. &
\textbf{\(d_a\):} A detailed account from the postmortem identifies late courier handoffs as the reason for customer refund requests and rejects the reverse explanation. \\
\cmidrule(lr){2-3}
&
\textbf{BA:} Find the account stating that customer refund requests caused late courier handoffs in an online retailer's holiday postmortem. &
\textbf{\(d_b\):} A detailed account from the postmortem identifies customer refund requests as the reason for late courier handoffs and rejects the reverse explanation. \\
\bottomrule
\end{tabular}
\label{tab:switchir-realistic-example}
\end{table*}

\section{Full Realistic-Set Results}
\label{sec:full-realistic-results}

Table~\ref{tab:full-results-realistic} reports all four metrics for every category on the realistic set. The main-text summary retains the category-specific primary joint metrics.

\begin{table*}[!tbp]
\caption{Full zero-shot results on the realistic set. PAcc, FBoth, PAll, and CAll abbreviate PairAcc, FlipBoth, ParaAll, and ContrastAll. Random values are expectations under independent uniform selection; rerankers are reported separately.}
\centering
\footnotesize
\setlength{\tabcolsep}{1.2pt}
\begin{tabular}{@{}llcccccccccc@{}}
\toprule
Type & Model & \multicolumn{2}{c}{Constraint Shift} & \multicolumn{4}{c}{Explicit Filtering} & \multicolumn{4}{c}{Relational Flip} \\
\cmidrule(lr){3-4}\cmidrule(lr){5-8}\cmidrule(lr){9-12}
& & PAcc & FBoth & PAcc & FBoth & PAll & CAll & PAcc & FBoth & PAll & CAll \\
\midrule
Random & Random
& \scorecell{50}{0.500} & \scorecell{25}{0.250}
& \scorecell{50}{0.500} & \scorecell{25}{0.250} & \scorecell{25}{0.250} & \scorecell{6}{0.063}
& \scorecell{50}{0.500} & \scorecell{25}{0.250} & \scorecell{25}{0.250} & \scorecell{6}{0.063} \\
\midrule
Lexical & BM25
& \scorecell{0}{0.000} & \scorecell{0}{0.000}
& \scorecell{0}{0.000} & \scorecell{0}{0.000} & \scorecell{0}{0.000} & \scorecell{0}{0.000}
& \scorecell{3}{0.030} & \scorecell{0}{0.005} & \scorecell{2}{0.025} & \scorecell{0}{0.000} \\
\midrule
Retriever & Qwen3-Emb-0.6B & \scorecell{92}{0.925} & \scorecell{85}{0.850} & \scorecell{82}{0.825} & \scorecell{65}{0.650} & \scorecell{82}{0.825} & \scorecell{65}{0.650} & \scorecell{52}{0.518} & \scorecell{14}{0.135} & \scorecell{41}{0.410} & \scorecell{2}{0.020} \\
& Qwen3-Emb-4B & \scorecell{100}{1.000} & \scorecell{100}{1.000} & \scorecell{95}{0.950} & \scorecell{90}{0.900} & \scorecell{92}{0.925} & \scorecell{85}{0.850} & \scorecell{60}{0.600} & \scorecell{21}{0.210} & \scorecell{54}{0.540} & \scorecell{10}{0.100} \\
& Qwen3-Emb-8B & \scorecell{100}{1.000} & \scorecell{100}{1.000} & \scorecell{91}{0.913} & \scorecell{82}{0.825} & \scorecell{88}{0.875} & \scorecell{75}{0.750} & \scorecell{62}{0.618} & \scorecell{24}{0.240} & \scorecell{55}{0.550} & \scorecell{12}{0.120} \\
& E5-Mistral & \scorecell{98}{0.975} & \scorecell{95}{0.950} & \scorecell{74}{0.738} & \scorecell{48}{0.475} & \scorecell{72}{0.725} & \scorecell{45}{0.450} & \scorecell{61}{0.610} & \scorecell{24}{0.235} & \scorecell{54}{0.535} & \scorecell{12}{0.120} \\
& GritLM & \scorecell{100}{1.000} & \scorecell{100}{1.000} & \scorecell{96}{0.963} & \scorecell{92}{0.925} & \scorecell{95}{0.950} & \scorecell{90}{0.900} & \scorecell{66}{0.663} & \scorecell{32}{0.325} & \scorecell{57}{0.575} & \scorecell{18}{0.180} \\
& Promptriever & \scorecell{98}{0.975} & \scorecell{95}{0.950} & \scorecell{100}{1.000} & \scorecell{100}{1.000} & \scorecell{100}{1.000} & \scorecell{100}{1.000} & \scorecell{56}{0.558} & \scorecell{17}{0.170} & \scorecell{44}{0.445} & \scorecell{1}{0.010} \\
\midrule
Late interaction & BGE-M3-ColBERT & \scorecell{92}{0.925} & \scorecell{85}{0.850} & \scorecell{71}{0.713} & \scorecell{45}{0.450} & \scorecell{55}{0.550} & \scorecell{15}{0.150} & \scorecell{59}{0.590} & \scorecell{25}{0.250} & \scorecell{44}{0.435} & \scorecell{4}{0.040} \\
\midrule
Reranker & Qwen3-Rerank-0.6B & \scorecell{72}{0.725} & \scorecell{65}{0.650} & \scorecell{100}{1.000} & \scorecell{100}{1.000} & \scorecell{100}{1.000} & \scorecell{100}{1.000} & \scorecell{49}{0.493} & \scorecell{27}{0.270} & \scorecell{28}{0.285} & \scorecell{9}{0.090} \\
& Qwen3-Rerank-4B & \scorecell{100}{1.000} & \scorecell{100}{1.000} & \scorecell{100}{1.000} & \scorecell{100}{1.000} & \scorecell{100}{1.000} & \scorecell{100}{1.000} & \scorecell{82}{0.820} & \scorecell{67}{0.670} & \scorecell{70}{0.705} & \scorecell{47}{0.470} \\
\bottomrule
\end{tabular}
\label{tab:full-results-realistic}
\end{table*}


\section{Model-Specific Instruction Templates}
\label{sec:model-instruction-templates}

Table~\ref{tab:model-instruction-templates} records the exact model-side
formatting used in our implementation. Bracketed terms denote fields supplied
by SwitchIR. Documents are encoded without an instruction for the embedding
and late-interaction retrievers.

\begin{table*}[!t]
\centering
\footnotesize
\caption{Model-specific instruction templates used in the experiments. Models
on the same row share the same implementation. Line breaks shown in the
templates are preserved in the model input.}
\label{tab:model-instruction-templates}
\setlength{\tabcolsep}{5pt}
\renewcommand{\arraystretch}{1.15}
\begin{tabular}{@{}p{0.21\textwidth}p{0.73\textwidth}@{}}
\toprule
Model & Model input \\
\midrule
Qwen3-Embedding
(0.6B, 4B, 8B);
E5-Mistral
&
Instruction input: [instruction]\newline
Query input:[query]
\\
\midrule
GritLM
&
Instruction input: <|user|> [instruction] <|embed|> \newline
Query input:[query]
\\
\midrule
Promptriever
&
Query input:
query:\ \ [query] [instruction]
\newline
Document input:
passage:\ \ [document].
\\
\midrule
BGE-M3-ColBERT
&
Query input:
[instruction] [query] 
\newline
Document input:
[document].
\\
\midrule
Qwen3-Reranker
(0.6B, 4B)
&
Instruction input: [instruction] \newline
Query input:[query]\newline
Document input : [document]
\\
\bottomrule
\end{tabular}
\end{table*}

BM25 has no model-specific instruction interface and is therefore not included
in Table~\ref{tab:model-instruction-templates}.

\section{Input-Format Ablation}
\label{sec:input-ablation-appendix}

Table~\ref{tab:input-ablation} averages the seven non-reranker neural retrieval runs after alias deduplication. The joint metric is named explicitly for each category. Query-only inputs yield zero joint accuracy because the same query cannot resolve opposing preferences. Concatenation and separate instructions recover high Constraint Shift and Explicit Filtering scores. Relational Flip ContrastAll remains low at \(0.007\) versus \(0.015\) controlled and \(0.061\) versus \(0.084\) realistic, with a modest advantage for native instruction fields.

\begin{table}[!tbp]
\centering
\small
\caption{Input-format ablation, macro-averaged over seven non-reranker neural retrieval runs after alias deduplication. Controlled and realistic results are separate panels.}
\setlength{\tabcolsep}{3pt}
\begin{tabular}{lccc}
\toprule
Input format & Constraint & Explicit & Relational \\
& FlipBoth & ContrastAll & ContrastAll \\
\midrule
\multicolumn{4}{l}{\textit{Controlled set}} \\
Query only & 0.000 & 0.000 & 0.000 \\
Instruction + query & 0.980 & 0.871 & 0.007 \\
Separate instruction & 0.986 & 0.923 & 0.015 \\
\midrule
\multicolumn{4}{l}{\textit{Realistic set}} \\
Query only & 0.000 & 0.000 & 0.000 \\
Instruction + query & 0.943 & 0.650 & 0.061 \\
Separate instruction & 0.943 & 0.679 & 0.084 \\
\bottomrule
\end{tabular}
\label{tab:input-ablation}
\end{table}

\section{Diagnostic Training Details}
\label{sec:training-details}

\begin{table}[!ht]
\centering
\footnotesize
\caption{Reproduction configuration for both diagnostic-training experiments.}
\label{tab:training-config}
\begin{tabular}{@{}p{0.22\columnwidth}p{0.72\columnwidth}@{}}
\toprule
Setting & Configuration \\
\midrule
Adaptation &
Qwen3-Embedding-0.6B and 4B; LoRA rank 8, \(\alpha=16\), dropout 0.05, applied to query, key, value, and output projections. \\
Training and reporting &
Controlled data only; AdamW, learning rate \(2\times10^{-4}\), batch size 8, at most 5 epochs, early-stopping patience 2, and \(\tau=0.02\). Checkpoints use controlled-dev FlipBoth with PairAcc as tie-break. Realistic data are external evaluation only. Seeds are 42--46; results use equal-weight relation-family macro-averaging followed by the across-seed mean and sample standard deviation. \\
\bottomrule
\end{tabular}
\end{table}

\begin{figure}[!ht]
\centering
\includegraphics[width=\columnwidth]{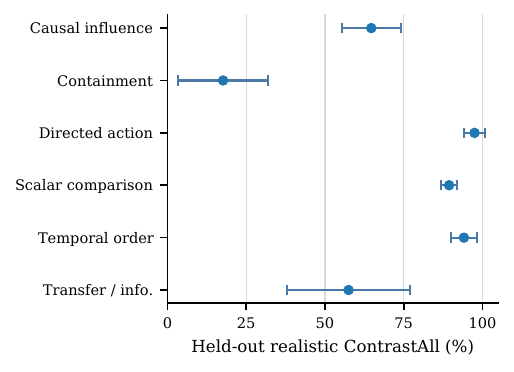}
\caption{Relation-family heterogeneity for Qwen3-Embedding-4B on realistic leave-one-relation-out evaluation. Points show mean ContrastAll over five seeds and error bars show sample standard deviation.}
\label{fig:training-heterogeneity}
\end{figure}

Figure~\ref{fig:training-heterogeneity} shows that held-out transfer is strongly relation dependent. ContrastAll ranges from \(17.6\pm14.4\) for Containment/Hierarchy to \(97.5\pm3.4\) for Directed Action; Scalar Comparison and Temporal Order also exceed \(89\%\), while Causal Influence and Transfer/Information Flow reach \(64.7\%\) and \(57.5\%\). The \(70.1\%\) realistic macro-average therefore summarizes a broad range of relation-specific outcomes.

\end{document}